\documentclass[twoside]{article}
\usepackage{qic,epsfig}
\usepackage{float}

\def\be{\begin{eqnarray}}
\def\ee{\end{eqnarray}}
\begin{document}
\setlength{\textheight}{8.0truein}    %FOR 2ND PAGE ONWARDS

\runninghead{Optimal Bacon-Shor Codes}
            {J. Napp and J. Preskill}

\normalsize\textlineskip
\thispagestyle{empty}
\setcounter{page}{1}
%%% CALT 68-2887
%%% Hide copyright in preprint
%\copyrightheading{Vol.}{No.}{Year}{Page Nos.}
%%%

\vspace*{0.88truein}

\alphfootnote

\fpage{1}

\centerline{\bf
%%%%%%%%%%%%%%%%%%%%%
%Put in titles here
%%%%%%%%%%%%%%%%%%%%%
OPTIMAL BACON-SHOR CODES}
%\centerline{\bf }
\vspace*{0.37truein}
\centerline{\footnotesize
%%%%%%%%%%%%%%%%%%%%%%%%%%%%%%%%%%%%
%put authors' name and address here
%%%%%%%%%%%%%%%%%%%%%%%%%%%%%%%%%%%%
JOHN NAPP and JOHN PRESKILL}
\vspace*{0.015truein}
\centerline{\footnotesize\it Institute for Quantum Information and Matter, California Institute of Technology }
\baselineskip=10pt
\centerline{\footnotesize\it Pasadena, CA 91125, USA}
\vspace*{0.225truein}

%%% Hide dates in preprint
%\publisher{(received date)}{(revised date)}
%%%
%%% Date of draft
%\publisher{DRAFT}{4 September 2012}
%
\vspace*{0.21truein}
%
%% \abstracts{first paragraph}{second paragraph}{third paragraph}
%% If there is only one paragraph, just keep the second and third empty 
%% like the following one 
\abstracts{
%%%%%%%%%%%%%%%%%%%%
% put abstract here
%%%%%%%%%%%%%%%%%%%%
%%
We study the performance of Bacon-Shor codes, quantum subsystem codes which are well suited for applications to fault-tolerant quantum memory because the error syndrome can be extracted by performing two-qubit measurements. Assuming independent noise, we find the optimal block size in terms of the bit-flip error probability $p_X$ and the phase error probability $p_Z$, and determine how the probability of a logical error depends on $p_X$ and $p_Z$. We show that a single Bacon-Shor code block, used by itself without concatenation, can provide very effective protection against logical errors if the noise is highly biased ($p_Z/p_X\gg 1)$ and the physical error rate $p_Z$ is a few percent or below. We also derive an upper bound on the logical error rate for the case where the syndrome data is noisy. 
}{}{}
\vspace*{10pt}

\vspace*{3pt}
%%% Hide communicate in preprint
%\communicate{to be filled by the Editorial}
%%%
%%
%%
\vspace*{1pt}\textlineskip    %) USE THIS MEASUREMENT WHEN THERE IS
   %) A SECTION HEADING
%\vspace*{-0.5pt}
%\noindent
%%%%%%%%%%%%%%%%%%%%%%%%%%%%%%%%
%put the text of the paper here
%%%%%%%%%%%%%%%%%%%%%%%%%%%%%%%%%\documentclass[aps,pra,showpacs,preprint]{revtex4}

\section{Introduction}
\label{sec:intro}
Bacon-Shor codes \cite{Bacon2006,Shor1995} are quantum subsystem codes \cite{Kribs2006,Poulin2005} which are well suited for applications to fault-tolerant quantum memory \cite{Shor1996,Gottesman2009}, because error syndrome information can be extracted by measuring only two-qubit operators that are spatially local if the qubits are arranged in a two-dimensional lattice. In this paper we assess the performance of these codes.

We consider noise models such that qubits in the code block are subject to both bit flip ($X$) errors and dephasing ($Z$) errors, where the bit flips occur with probability $p_X$ and phase errors occur with probability $p_Z$. We assume that the noise acts independently on each qubit, and that the $X$ and $Z$ errors are uncorrelated. Under these assumptions we find the optimal block size of the code, and the failure probability achieved by this optimal code. We obtain analytic formulas for the optimal failure probability for the case of unbiased noise ($p_X = p_Z$) and the case of highly biased noise ($b\equiv p_Z/p_X \gg 1$). In both cases the formula applies in the asymptotic limit of small $p_Z$. Our results show that the failure probability of the optimal code falls exponentially in $1/p_Z$. We also consider the case where the code's syndrome bits are prone to error, and derive upper bounds on the failure probability in that case. 

We find that for the case of unbiased noise ($p_Z=p_X \equiv p$), the optimal Bacon-Shor code achieves the failure probability
\begin{eqnarray}\label{eq:symmetric-optimal-intro}
BSOptimalFailProb(p)&=& \left(\frac{2}{\pi\ln 2}\right)^{1/2}\exp\left(\frac{\ln^2 2}{8}\right) p^{1/2}\exp\left(-\frac{\ln^2 2}{8p}+O(p)\right)\nonumber\\
\end{eqnarray}
or 
\begin{eqnarray}
\ln[BSOptimalFailProb(p)]&=& -A/p -(0.5)\ln(1/p) +C +O(p),
\end{eqnarray}
where
\begin{eqnarray}
A= .0600566,\quad C= .0175217.
\end{eqnarray}
For the case of highly biased noise, we find
\begin{eqnarray}
\ln[BSOptimalFailProb(p_Z,b)] = -A(b)/p_Z -(0.5)\ln(1/p_Z) +C(b)+  O(p_Z ~{\rm polylog}(b)), %+ O(p_Z\ln^4 b)
\end{eqnarray}
where 
\begin{eqnarray}
A(b) = \frac{1}{8}\left(W(\sqrt{b})\right)^2 +O(b^{-1/2}\ln b);
\end{eqnarray}
here $W$ denotes the Lambert $W$ function, with asymptotic expansion
\begin{eqnarray}
W(\sqrt{b}) = \ln \sqrt{b} - \ln\ln \sqrt{b} + \frac{\ln\ln\sqrt{b}}{\ln\sqrt{b}} + O\left(\frac{\ln\ln\sqrt{b}}{\ln^2\sqrt{b}}\right).
\end{eqnarray}
We also compute the asymptotic form of $C(b)$ for $b \gg 1$.

In the case where the error syndrome is noisy, the reliability of the syndrome can be improved by measuring it repeatedly. We consider an idealized noise model such that qubit errors and syndrome measurement errors are equally likely and independent. For that model we derive a lower bound on the coefficient $\tilde A(b)$ of $1/p_Z$ in the natural logarithm of the logical failure rate, finding
\begin{eqnarray}
\tilde A(b) \ge \frac{Z(b)}{4\mu^2} W\left(\sqrt{\frac{b}{4Z(b)}}~\right),
\end{eqnarray}
where $\mu = 2.63816$ is the connective constant of a self-avoiding walk on a two-dimensional square lattice, and $Z(b)$ is a slowly varying monotonic function which ranges from $1/e$ to 1 as $b$ increases from 1 to infinity. This bound applies for any value of $b\ge 1$. 

Previous work \cite{AliferisCross2007,Cross2007} on the applications of Bacon-Shor codes to fault-tolerant quantum computing has focused on relatively small codes used at the bottom layer of a concatenated coding scheme. Here we emphasize that a sufficiently large Bacon-Shor code, used by itself without concatenation, can also be quite effective if the physical error rate is low enough. For example, if the syndrome is perfect, then the probability of a logical failure is below $2\times 10^{-19}$ for $p_Z = .01$, $b=100$, and below $10^{-12}$ for $p_Z= .03$, $b=1000$. Fault-tolerant circuits based on Bacon-Shor codes will be more fully discussed and analyzed in a separate paper \cite{Brooks2012}, where we consider in particular the case of highly biased noise. 

\section{2D Bacon-Shor code}

The two-dimensional Bacon-Shor code \cite{Bacon2006,Shor1995} is a quantum subsystem code (really a family of codes), which pieces together two dual quantum repetition codes, one protecting against bit flip errors and one protecting against phase errors.  Like any subsystem code \cite{Kribs2006,Poulin2005}, it can be defined by its ``gauge algebra'' --- a set of Pauli operators that commute with the logical operators. The center of the gauge algebra is the code's stabilizer group, and the code space is determined by fixing the eigenvalues of all the stabilizer operators to be (say) $+1$. Other operators in the gauge algebra act nontrivially on ``gauge qubits'' but trivially on the code's protected qubits.

Specifically, we consider combining a length-$m$ repetition code to protect against $Z$ errors with a length-$n$ repetition code to protect against $X$ errors, where $Z$ and $X$ denote the single-qubit Pauli operators $\sigma_Z$ and $\sigma_X$. The resulting Bacon-Shor code block contains $mn$ qubits arranged at the sites of an $m \times n$ square lattice. The gauge algebra is generated by two-qubit operators $XX$ acting on pairs of neighboring qubits in the same horizontal row of the lattice and two-qubit operators $ZZ$ acting on neighboring qubits in the same vertical column. There is one protected qubit, and we may choose the logical Pauli operators to be $\bar X = X^{\otimes n}$ acting on all the qubits in a column and $\bar Z = Z^{\otimes m}$ acting on all the qubits in a row. The stabilizer group has $m+n-2$ generators --- there are $m-1$ $X$-type generators, each a product of $2n$ $X$s acting on all the qubits in a pair of neighboring columns, and $n-1$ $Z$-type generators, each a product of $2m$ $Z$s acting on all the qubits in a pair of neighboring rows. (We say that a Pauli operator is $X$-type if it is a tensor product of $X$s and identity operators, and we say it is $Z$-type if it is a tensor product of $Z$s and identity operators.) Aside from the protected logical system, there are $(m-1)(n-1)$ gauge qubits.

\begin{figure}[h!]
\begin{center}
\epsfig{file=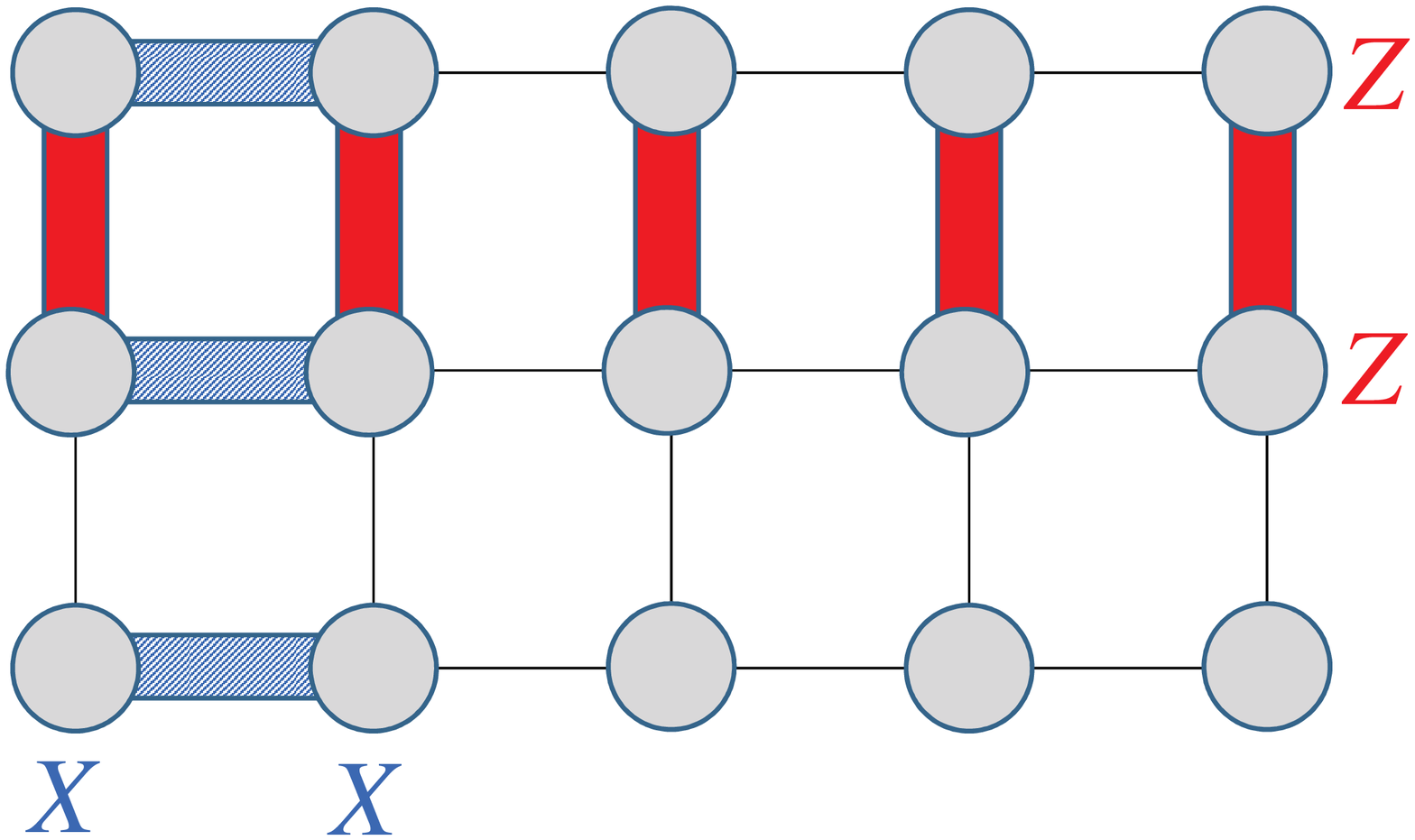,width=8cm} \vspace{0.2cm}
\end{center}
\fcaption{\label{fig:BS-block} Bacon-Shor code block for $m=5$ and $n=3$. A weight-($2m$) $Z$-type stabilizer generator is the product of $m$ weight-two $Z$-type gauge operators (solid red); a weight-($2n$) $X$-type stabilizer generator is the product of $n$ weight-two $X$-type gauge operators (hashed blue).}
\end{figure}

Though each $Z$-type stabilizer generator (or ``check operators'') has weight $2m$, its value can be ascertained by measuring $m$ weight-two gauge operators. For example, the product of $2m$ $Z$s in the first two rows is equivalent to the product of $m$ $Z$-type gauge operators, each acting on a pair of qubits in the first two positions of the same column. To determine the check operator, we may measure these $m$ gauge operators (obtaining either $+1$ or $-1$ for each outcome) and multiply the results. Similarly, the value of a weight-$2n$ $X$-type check operator can be found by measuring $n$ weight-two $X$-type gauge operators and multiplying the results. Furthermore, since all check operators commute with all gauge-qubit operators, measuring $X$-type gauge qubit operators to determine the value of an $X$-type check operator does not disturb the values of any $Z$-type check operators (though it may flip the value of $Z$-type gauge operators), and vice-versa.

If $m$ and $n$ are both odd, then the code can correct $(m-1)/2$ $Z$ errors and $(n-1)/2$ $X$ errors. To describe the error recovery procedure, it is useful to exploit the freedom to ``fix the gauge'', {\em i.e.} apply gauge operators to the block that commute with the logical qubit operators. An even number of $X$ errors in a row can be ``gauged away'' by applying gauge operators. If there are an odd number of $X$ errors in a row, we may ``gauge away'' all errors except one, which can be moved into the first position in the row. Thus, in an appropriate gauge, $X$ errors occur only in the first column. Likewise, in an appropriate gauge, $Z$ errors occur only in the first row. In this gauge, then, the check operators test whether two neighboring qubits in the first column agree in the $Z$ basis, and whether two neighboring qubits in the first row agree in the $X$ basis. Error recovery proceeds by applying $X$ to at most $(n-1)/2$ qubits in the first column, to restore all $Z$-type check operators to the value $+1$, and by applying $Z$ to at most $(m-1)/2$ qubits in the first row, to restore all $X$-type check operators to the value $+1$. Thus, error recovery is successful if the number of rows with an odd number of $X$ errors is at most $(n-1)/2$ and if the number of columns with an odd number of $Z$ errors is at most $(m-1)/2$.

If we choose the gauge so that $ZZ=+1$ for any pair of qubits in the same column, then the eigenstates of $\bar X$ with eigenvalues $\pm 1$ become
\begin{equation}
|\pm\rangle_C \propto \left(|00\cdots 0\rangle \pm |11\cdots 1\rangle\right)^{\otimes m};
\end{equation}
these are tensor products of $m$ length-$n$ ``cat states'' in the standard basis, one for each column. If we choose the gauge so that $XX=+1$ for any pair of qubits in the same row, then the eigenstates of $\bar Z$ with eigenvalues $\pm 1$ becomes
\begin{eqnarray}
|0\rangle_C &\propto& \left(|++\cdots +\rangle + |--\cdots -\rangle\right)^{\otimes n},\nonumber\\
|1\rangle_C &\propto& \left(|++\cdots +\rangle - |--\cdots -\rangle\right)^{\otimes n};
\end{eqnarray}
these are tensor products of $n$ length-$m$ ``cat states'' in the dual basis, one for each row. 

To perform a destructive measurement of the logical operator $\bar Z$, we can measure all $mn$ qubits in the $Z$ basis, compute the parity of the outcomes for each row, and then decode the measurement by performing a majority vote on the row parities. Likewise, to perform a destructive measurement of the logical operator $\bar X$, we can measure all $mn$ qubits in the $X$ basis, compute the parity of the outcomes for each column, and then decode the measurement by performing a majority vote on the column parities. 

\section{Failure probability for the 2D Bacon-Shor code}

Consider an independent noise model in which the probability of an $X$ error for each qubit is $p_X$, and the probability of a $Z$ error is $p_Z$ (that is, for each qubit, there is no error with probability $(1-p_X)(1-p_Z)$, an $X$ error with probability $p_X(1-p_Z)$, a $Z$ error with probability $(1-p_X)p_Z$ and an $iY=ZX$ error with probability $p_Xp_Z$). For any fixed values of $p_X$ and $p_Z$, there is an optimal choice for the code dimensions $m$ and $n$ which minimizes the probability of a logical error. We will estimate the optimal failure probability in the limit of small $p$. Since $X$ and $Z$ error recovery are identical (except for the interchange of the rows and columns of the lattice), we will estimate the probability of a logical $Z$ error; the same derivation also applies to the probability of a logical $X$ error. 

First we note that for a column of length $n$, if $Z$ errors occur independently with probability $p$ at each position, then the probability of an odd number of errors in the column is
\begin{equation}
OddProb(p,n) = \frac{1}{2}\left(1 - (1-2p)^n\right).
\end{equation}
We derive this formula by observing that the terms of even order in $p$ in  the binomial expansions of $\left( (1-p) + p \right)$ and $\left( (1-p) - p \right)$ are identical, while the terms of odd order in $p$ are the same except for a sign flip; hence the difference $\frac{1}{2}\left(1 - (1-2p)^n\right)$ sums up all the odd-order terms. To minimize the probability of a logical error for given $p$, we will choose $n$ and $m$  so that $p^2mn$ is O(1). Therefore in the limit of large $n$ we may use the approximation
\begin{equation}
OddProb(p,n) = \frac{1}{2}\left(1 - \exp\left[-2p(1+p)n + O(p^3n)\right]\right).
\end{equation}
We keep track of the $O(p^2n)$ in the exponential because $OddProb(p,n)$ will be raised to a power $O(m)$ in the expression for the Bacon-Shor failure probability.

Now consider a length-$m$ repetition code, where bit errors occur independently with probability $x$. If $m$ is odd, an encoded error occurs if the number of bit errors is $(m+1)/2$ or more. Therefore the failure probability of the repetition code is
\begin{eqnarray}
RepFailProb(x,m) &=&\sum_{k=(m+1)/2}^m {m\choose k}x^k (1-x)^{m-k} \nonumber\\
&=& \left(\frac{x}{1-x}\right)^{1/2}\left[x(1-x)\right]^{m/2}\sum_{r=0}^{(m-1)/2}{m\choose \frac{m+1}{2}+r}\left(\frac{x}{1-x}\right)^{r}.
\end{eqnarray}
From the Stirling approximation,
\begin{equation}
{m\choose \frac{m+1}{2}+r}\approx 2^m \left(\frac{2}{\pi m}\right)^{1/2}\exp\left(-\frac{2}{m}\left(r+\frac{1}{2}\right)^2\right),
\end{equation}
neglecting a multiplicative $O(1/m)$ correction. Making another $O(1/m)$ multiplicative error, we may replace the exponential inside the sum over $r$ by 1, obtaining
\begin{equation}
RepFailProb(x,m) \approx 2^m \left(\frac{2}{\pi m}\right)^{1/2}\left(\frac{x}{1-x}\right)^{1/2}\left[x(1-x)\right]^{m/2}\sum_{r=0}^{(m-1)/2}\left(\frac{x}{1-x}\right)^{r},
\end{equation}
and we also make a negligible error by extending the upper limit on the sum to infinity, finding
\begin{equation}
\sum_{r=0}^{\infty}\left(\frac{x}{1-x}\right)^{r}= \frac{1-x}{1-2x},
\end{equation}
and thus
\begin{equation}
RepFailProb(x,m) \approx \left(\frac{2}{\pi m}\right)^{1/2}\left(\frac{x(1-x)}{(1-2x)^2}\right)^{1/2}\left[4x(1-x)\right]^{m/2}.
\end{equation}

To compute the probability of a $Z$-type logical error for the Bacon-Shor code, we substitute $OddProb(p,n)$ for $x$ in the expression for $RepFailProb(x,m)$, finding
\begin{eqnarray}\label{eq:order-p-correction}
\left[4x(1-x)\right]^{m/2} &\approx& \left( 1- \exp\left[-4p(1+p)n+O(p^3n)\right]\right)^{m/2}\nonumber\\
&=& \exp\left(\frac{2p^2mn[1+O(p)+O(p^2n)]}{e^{4pn}-1}\right)\left(1-e^{-4pn}\right)^{m/2}, 
\end{eqnarray}
and
\begin{eqnarray}
\frac{x(1-x)}{(1-2x)^2}\approx \frac{1}{4}\left(e^{4pn}-1\right);
\end{eqnarray}
hence
\begin{eqnarray}\label{eq:prob-Z}
BSZFailProb(p_Z,m,n)&=& RepFailProb(OddProb(p_Z,n),m)\nonumber\\
&\approx& \frac{1}{\sqrt{2\pi m}}\exp\left[\frac{2p_Z^2mn}{e^{4p_Zn}-1}\right]\left(e^{4p_Zn}-1\right)^{1/2}\left(1-e^{-4p_Zn}\right)^{m/2},\nonumber\\
\end{eqnarray}
up to multiplicative corrections higher order in $p_Z$ (assuming $p_Zn$ and $p_Z m$ are O(1)). By the same argument, the probability of an $X$-type logical error is given by a similar expression, but with $m$ and $n$ interchanged and $p_Z$ replaced by $p_X$:
\begin{eqnarray}\label{eq:prob-X}
BSXFailProb(p_X,m,n)&=& RepFailProb(OddProb(p_X,m),n) \nonumber\\
&\approx& \frac{1}{\sqrt{2\pi n}}\exp\left[\frac{2p_X^2mn}{e^{4p_Xm}-1}\right]\left(e^{4p_Xm}-1\right)^{1/2}\left(1-e^{-4p_Xm}\right)^{n/2},\nonumber\\
\end{eqnarray}
up to multiplicative corrections higher order in $p_X$ (assuming $p_Xm$ and $p_X n$ are O(1)).

\subsection{Unbiased noise}

If the noise is unbiased ($p_X = p_Z\equiv p$), then the optimal Bacon-Shor code is symmetric ($m=n$), and the $X$-type and $Z$-type logical errors occur with equal probability.
To find the optimal value of $n$ for small $p$, we note that
\begin{eqnarray}
\ln\left[BSFailProb(p,n)\right]\approx \frac{n}{2}\ln\left(1-e^{-4pn}\right)+\cdots,
\end{eqnarray}
where the ellipsis indicates corrections suppressed by powers of $p$ or $1/n$. This expression, regarded as a function of $n$, attains its minimum when $y=pn$ satisfies
\begin{equation}
\ln\left(1- e^{-4y}\right) + \frac{4y}{e^{4y}-1} = 0,
\end{equation}
or $4y = \ln 2$. (Though $n$ is actually required to be an odd integer, we may ignore this requirement if $p$ is small and $n$ is correspondingly large.) Substituting $4pn\approx \ln 2$ into the expression for the Bacon-Shor failure probability Eq.(\ref{eq:prob-Z}) we find
\begin{eqnarray}\label{eq:symmetric-optimal}
BSOptimalFailProb(p)&\approx& \left(\frac{2}{\pi\ln 2}\right)^{1/2}\exp\left(\frac{\ln^2 2}{8}\right) p^{1/2}\exp\left(-\frac{\ln^2 2}{8p}\right)\nonumber\\
&=& (1.01768) \sqrt{p}\exp\left(-\frac{.0600566}{p}\right), 
\end{eqnarray}
or 
\begin{eqnarray}
\ln[BSOptimalFailProb(p)]&\approx& -A/p +B\ln p +C +O(p)
\end{eqnarray}
where
\begin{eqnarray}
A= .0600566,\quad B= .5, \quad C= .0175217.
\end{eqnarray}
%I have also estimated the O($p$) correction as $-(10.08) p +O(p^2)$. 
This optimal failure probability is achieved by choosing the linear size $n$ of the code such that
\begin{equation}
pn = \frac{1}{4}\ln 2 = .173287.
\end{equation}
In terms of the asymptotically optimal linear size $n=\frac{\ln 2}{4p}$, the optimal failure probability can be expressed as
\begin{eqnarray}
BSOptimalFailProb(n)&\approx& \left(\frac{2}{\pi\ln 2}\right)^{1/2}\exp\left(\frac{\ln^2 2}{8}\right) \left(\frac{\ln 2}{4n}\right)^{1/2}\exp\left(-\frac{n \ln 2}{2}\right)\nonumber\\
&=& \frac{2^{\ln 2/8}}{\sqrt{2\pi n}}~ ~2^{-n/2}.
\end{eqnarray}

%For example, for $p=.01$, the optimal value of $L$ is 17, and the corresponding failure probability is $2.307 \times 10^{-4}$, while our asymptotic formula (expressed as a function of $p$, without the correction linear in $p$ in the exponent) predicts $2.508\times 10^{-4}$. For $p=.003$, the optimal value of $L$ is $57$, and the corresponding failure probability is $1.096\times 10^{-10}$, while the asymptotic formula predicts $1.127 \times 10^{-10}$. For $p=.001$, the optimal value of $L$ is $173$, and the corresponding failure probability is $2.638\times 10^{-28}$, while the asymptotic formula predicts $2.663 \times 10^{-28}$. (Including the term linear in $p$ in the exponent improves the predictions to $2.268 \times 10^{-4}$, $1.094\times 10^{-10}$, $2.636 \times 10^{-28}$.) Although there is strictly speaking no accuracy threshold for this family of codes, the performance of the optimal code is very good when the error rate is sufficiently small. 

Although there is strictly speaking no accuracy threshold for this family of codes, the performance of the optimal code is very good when the error rate is sufficiently small. For example, for $p=.001$, the optimal value of $n$ is $173$, and the corresponding failure probability is $2.638\times 10^{-28}$, while the asymptotic formula Eq.(\ref{eq:symmetric-optimal}) predicts $2.663 \times 10^{-28}$.

\subsection{Highly biased noise}

If the noise is highly biased ($b \equiv p_Z/p_X \gg 1$), then the optimal Bacon-Shor code is  highly asymmetric --- it combines a length-$m$ repetition code to protect against $Z$ errors with a length-$n$ repetition code to protect against $X$ errors, where $m \gg n$. In the limit of large noise bias, we can find analytic formulas for the optimal values of $m$ and $n$ and for the corresponding optimal probability of a logical error.

From Eq.(\ref{eq:prob-Z}) and Eq.(\ref{eq:prob-X}), we have 
\begin{eqnarray}\label{eq:XZ-log}
\ln\left[BSZFailProb(p_Z,m,n)\right]&\approx& \frac{m}{2}\ln\left(1-e^{-4p_Zn}\right)+\cdots,\nonumber\\
\ln\left[BSXFailProb(p_X,m,n)\right]&\approx& \frac{n}{2}~\ln\left(1-e^{-4p_Xm}\right)+\cdots.
\end{eqnarray}
where the ellipsis indicates corrections suppressed by powers of $p_Z$, $p_X$, $m^{-1}$, $n^{-1}$. 
For $b\equiv p_Z/p_X \gg 1$, the optimal values of $m$ and $n$ are such that $4 p_X m$ and $e^{-4p_Zn}$ are both small; hence we can justify keeping the leading terms in a power series expansion in these small quantities, obtaining 
\begin{eqnarray}\label{eq:further-approx}
\ln\left[BSZFailProb(p_Z,m,n)\right]&\approx& -\frac{m}{2}\exp\left(-4p_Zn\right)-\frac{m}{4} \exp\left(-8p_Z n\right)+\cdots,\nonumber\\
\ln\left[BSXFailProb(p_X,m,n)\right]&\approx& \frac{n}{2}\ln\left(4p_Xm\right)-p_Xmn +\cdots.
\end{eqnarray}
For now we neglect the non-leading terms in the expansion of both logarithms, which we will verify  {\em a posteriori} are additive corrections of order $b^{-1/2}\ln b$. 

We will minimize the failure probability subject to the constraint that (the leading contributions to) the logs of the $Z$ and $X$ failure probabilities are equal. Because we are neglecting the higher order terms in both Eq.(\ref{eq:XZ-log}) and Eq.(\ref{eq:further-approx}), the solution we find many not be the true optimum, but we will see that it provides an upper bound on the optimal failure probability which is reasonably tight provided $p_Z \ln^4 b$ and $b^{-1/2}\ln b$ are small.
%this constraint would also be obtained asymptotically for large $b$ if we minimized the sum of the $Z$ and $X$ failure probabilities. %[{\em Is that true?}] 
Defining the variables 
\begin{eqnarray}
X = 4 p_X m, \quad Z = 4 p_Z n,
\end{eqnarray}
we want to find the values of $X$ and $Z$ that minimize the function
\begin{eqnarray}
F(X,Z) = Z\ln X
\end{eqnarray}
(the log of the failure probability multiplied by $8 p_Z = 8 b p_X$) subject to the constraint
\begin{eqnarray}\label{eq:constraint}
Z \ln X = - bX e^{-Z}.
\end{eqnarray}
Introducing a Lagrange multiplier $\lambda$ to do the constrained minimization, we obtain the equations
\begin{eqnarray}
-(\lambda -1) Z &=& \lambda bX e^{-Z}=(\lambda -1) \ln X 
\end{eqnarray}
together with the constraint equation Eq.(\ref{eq:constraint}); these equations imply
\begin{eqnarray}
X = e^{-Z} \quad {\rm and} \quad Z e^Z = \sqrt{b} \quad \Rightarrow \quad X = Z /\sqrt{b}.
\end{eqnarray}
Evaluating $F$ at its minimum yields
\begin{equation}
F_{\rm min} = -Z^2.
\end{equation}

The solution to the equation $Z e^Z = \sqrt{b}$ is the Lambert $W$ function $Z= W(\sqrt{b})$, which has the asymptotic expansion
\begin{eqnarray}
Z = W(\sqrt{b}) = \ln \sqrt{b} - \ln\ln \sqrt{b} + \frac{\ln\ln\sqrt{b}}{\ln\sqrt{b}} + O\left(\frac{\ln\ln\sqrt{b}}{\ln^2\sqrt{b}}\right)
\end{eqnarray}
for $b\gg 1$, and the log of the optimal failure probability is 
\begin{eqnarray}\label{eq:leading-optimal}
\ln\left[BSZFailProb(p_Z,m,n)\right]\approx \ln\left[BSXFailProb(p_Z,m,n)\right]\approx - \frac{W^2(\sqrt{b})}{8 p_Z} +\cdots.
\end{eqnarray}
The optimal code has dimensions
\begin{eqnarray}\label{eq:optimal-nm}
n &=& \frac{Z}{4p_Z}  \approx \frac{1}{4p_Z}  W(\sqrt{b})\approx \frac{1}{4p_Z}\ln \sqrt{b},\nonumber\\
m &=& \frac{X}{4p_X}  \approx \frac{1}{4p_Z} \sqrt{b}~ W(\sqrt{b})\approx \frac{1}{4p_Z} \sqrt{b} ~\ln \sqrt{b},
\end{eqnarray}
with aspect ratio $m/n= \sqrt{b}$. As the bias increases with $p_Z$ fixed, the code size creeps up slowly in the $X$-protection direction, and more rapidly in the $Z$-protection direction. 

For the optimal code the probability of an odd number of $X$ errors in a row decreases as $b$ increases according to
\begin{eqnarray}
OddProb(p_X,m) \approx \frac{1}{2}\left(1 - e^{-2p_Xm}\right)\approx p_X m \approx \frac{1}{4} X \approx  \frac{1}{4}\frac{W(\sqrt{b})}{\sqrt{b}}\approx \frac{1}{4} \frac{\ln\sqrt{b}}{\sqrt{b}},
\end{eqnarray}
while the probability of an odd number of $Z$ errors in a column asymptotically approaches $1/2$ according to
\begin{eqnarray}
OddProb(p_Z,n) \approx \frac{1}{2}\left(1 - e^{-2p_Zn}\right)=\frac{1}{2}\left(1 - e^{-Z/2}\right) \approx \frac{1}{2}\left(1 - \sqrt{X}\right),
\end{eqnarray}
or
\begin{eqnarray}
\frac{1}{2} - OddProb(p_Z,n) \approx \frac{1}{2}\left(\frac{W(\sqrt{b})}{\sqrt{b}}\right)^{1/2}.
\end{eqnarray}
%We see that even when $p_Z$ is relatively high and $p_X$ is relatively low it pays to devote some of the code's error-correcting power to controlling the $X$ errors, which would otherwise limit the code-length we could devote to controlling the $Z$ errors. 
The optimal failure probability is much smaller for $b\gg 1$ than in the case $b=1$ for the same value of $p_Z$, but the price paid is that the block size is correspondingly significantly larger:
\begin{eqnarray}
mn = \frac{\sqrt{b}~ W^2(\sqrt{b})}{16 p_Z^2},
\end{eqnarray}
as compared to 
\begin{eqnarray}
n^2 = \frac{\ln^2 2}{16 p^2}
\end{eqnarray}
in the unbiased case.

%We see that even when $p_Z$ is relatively high and $p_X$ is relatively low it pays to devote some of the code's error-correcting power to controlling the $X$ errors, which would otherwise limit the code-length we could devote to controlling the $Z$ errors. 

%To be concrete, consider the case $p_Z=.03$ and $b=1000$; using $W(\sqrt{b}) \approx 2.53$, the optimal code according to the above estimate has dimensions $ 666\times 21 $, with $Z$ oddness probability .3597 and $X$ oddness probability $.0200$. The actual optimal code has dimensions $ 721\times 21 $, with $Z$ oddness probability .3636 and $X$ oddness probability $.0212$.

Using Eq.(\ref{eq:optimal-nm}), we can now evaluate the subleading terms in Eq.(\ref{eq:further-approx}), finding
\begin{eqnarray}
-\frac{m}{4} \exp\left(-8p_Z n\right) &\approx& -\frac{m}{4}  X^2 \approx -\frac{1}{4}\left(\frac{\sqrt{b}~W(\sqrt{b})}{4p_Z}\right) \left(\frac{W(\sqrt{b})}{\sqrt{b}}\right)^2= -\frac{W^3(\sqrt{b})}{16 p_Z\sqrt{b}},
\nonumber\\
-p_Xmn  &\approx& -\frac{p_Z}{b}\left(\frac{\sqrt{b}~W(\sqrt{b})}{4p_Z}\right)\left(\frac{W(\sqrt{b})}{4p_Z}\right)= -\frac{W^2(\sqrt{b})}{16 p_Z\sqrt{b}}.
\end{eqnarray}
Aside from logarithmic factors, these terms are  suppressed by $O(b^{-1/2}\ln b)$ compared to the leading terms in Eq.(\ref{eq:further-approx}). We therefore expect our estimate for the optimal code dimensions to be accurate for $b \gg 1$. However, although these corrections are small compared to the leading terms, they are not necessarily negligible, and in particular they can become important as $p_Z\to 0$ with $b$ fixed. To obtain accurate results for small $p_Z$, we should resum the power series expansion in Eq.(\ref{eq:further-approx}), or in other words use the full expressions in Eq.(\ref{eq:XZ-log}), with the appropriate values of $m$, $n$, $Z$, and $X$ plugged in. Thus,
\begin{eqnarray}\label{eq:exponentiated-failure}
BSZOptimalFailProb(p_Z,b)&\approx&\left(1-e^{-Z}\right)^{m/2} \approx \left(1 - \frac{W(\sqrt{b})}{\sqrt{b}}\right)^{\frac{\sqrt{b}W(\sqrt{b})}{8p_Z}},\nonumber\\
BSXOptimalFailProb(p_Z,b)&\approx&\left(1-e^{-X}\right)^{n/2} \approx \left(1 - \exp\left(-e^{-W(\sqrt{b})}\right)\right)^{\frac{W(\sqrt{b})}{8p_Z}},
\end{eqnarray}
provide more accurate approximations than $\exp\left(- W^2(\sqrt{b})/8 p_Z\right)$. 

Plugging our solutions for $Z$ and $X$ back into the expressions for the prefactors in Eq.(\ref{eq:prob-Z}) and Eq.(\ref{eq:prob-X}), we again find somewhat different scaling with the bias for the two failure probabilities. For the $Z$ failure probability the prefactor is
\begin{eqnarray}
\frac{1}{\sqrt{2\pi m}}\exp\left[\frac{2p_Z^2mn}{e^{4p_Zn}-1}\right]\left(e^{4p_Zn}-1\right)^{1/2}
=\left(\frac{2 p_Z n}{\pi Zm}\right)^{1/2}\exp\left[\frac{\frac{1}{8} Z^2 \frac{m}{n}}{e^{Z}-1}\right]\left(e^{Z}-1\right)^{1/2};
\end{eqnarray}
making the approximation $e^Z -1 \approx e^Z$ as before, and using $m/n = \sqrt{b}=Ze^Z$, this becomes
\begin{eqnarray}\label{eq:Z-prefactor}
\approx\left(\frac{2 p_Z}{\pi}\right)^{1/2}b^{-1/4}Z^{-1/2}e^{ Z^3 /8}e^{Z/2}=\left(\frac{2 p_Z}{\pi}\right)^{1/2}Z^{-1}e^{ Z^3 /8}.
\end{eqnarray}
For the $X$ failure probability the prefactor is
\begin{eqnarray}
\frac{1}{\sqrt{2\pi n}}\exp\left[\frac{2p_X^2mn}{e^{4p_Xm}-1}\right]\left(e^{4p_Xm}-1\right)^{1/2}
= \left(\frac{2 p_Z }{\pi Z}\right)^{1/2}\exp\left[\frac{\frac{1}{8} X^2\frac{n}{m}}{e^{X}-1}\right]\left(e^{X}-1\right)^{1/2};
\end{eqnarray}
making the approximation $e^X-1 \approx X$ as before, this becomes
\begin{eqnarray}
\approx\left(\frac{2 p_Z}{\pi}\right)^{1/2}Z^{-1/2}\exp\left(\frac{ X }{8\sqrt{b}}\right) X^{1/2}=\left(\frac{2 p_Z}{\pi}\right)^{1/2}b^{-1/4}\exp\left(\frac{ X }{8\sqrt{b}}\right)
\approx \left(\frac{2 p_Z}{\pi}\right)^{1/2}b^{-1/4}.
\end{eqnarray}
We also note that, as in Eq.(\ref{eq:order-p-correction}), there are corrections to the factor $e^{Z^3/8}$ of the form
\begin{equation}
\exp\left(\frac{1}{8}Z^3\left[1 + O(p_Z) + O(p_Z Z)\right]\right) . %= e^{Z^3/8}\left(1 + O(p_ZZ^4)\right),
\end{equation}
%recalling $4p_Zn = Z$.
%Therefore, for Eq.(\ref{eq:Z-prefactor}) to be accurate we also require $p_Z Z^4\approx p_Z\ln^4(\sqrt{b}) \ll 1$. 
In fact we can sum up these corrections to all orders in $p_Z Z$ (still neglecting corrections suppressed by further powers of $p_Z$) to obtain the improved approximation
\begin{equation}
\exp\left(\frac{1}{8p}Z^2\left(1 - e^{-pZ}\right)\left[1 + O(p_Z)\right]\right).
\end{equation}
This correction can be significant for $pZ = O(1)$, but in that case our bound on the failure probability is rather loose anyway (see below), so including the correction is not so important.

Combining the estimated prefactors with Eq.(\ref{eq:exponentiated-failure}) we obtain the formulas for the failure probabilities:
\begin{eqnarray}\label{eq:including-prefactors}
BSZOptimalFailProb(p_Z,b)&\approx&\left(\frac{2 p_Z}{\pi}\right)^{1/2}Z^{-1}e^{ Z^3 /8}\left(1-\frac{Z}{\sqrt{b}}\right)^{\frac{Z\sqrt{b}}{8p_Z}},\nonumber\\
BSXOptimalFailProb(p_Z,b)&\approx& \left(\frac{2 p_Z}{\pi}\right)^{1/2}b^{-1/4}\left(1 - \exp\left(-e^{-Z}\right)\right)^{\frac{Z}{8p_Z}},
\end{eqnarray}
where $Z= W(\sqrt{b})$.
These asymptotic formulas apply if we fix the bias $b$ at a large value and then allow $p_Z$ to become sufficiently small. We see that the nonleading contribution to the log of the failure probability is approximately $\frac{1}{8} Z^3$ for $Z$-errors and approximately $- \frac{1}{4} \ln b$  for $X$ errors. These nonleading terms are small compared to the leading term $\frac{1}{8p_Z} Z^2$, which we minimized to obtain our estimates, provided $p_Z Z\approx p_Z \ln b \ll 1$. A numerical fit indicates that the leading correction to our asymptotic formula is approximately $(.01)p_Z (\ln b)^\delta$ where $\delta\approx 4$. 
%$p_Z \ln^\delta b$ where $\delta\approx 4$. 
When this correction is small, we expect our estimate of the optimal failure probability to be reasonably tight. 

%We may also ask: suppose the total number $q=mn$ of physical qubits in the code block is fixed. If the noise bias $b = p_Z/p_X$ is also fixed and we consider the limit of small $p_Z$, what is the optimal aspect ratio of the $m \times n$ rectangle, and what failure probability is achieved by this optimal code?

%In this case, we may consider both $p_Z n$ and $p_X m$ to be small, and neglect terms suppressed by powers of either quantity. Equating the logs of the $Z$ and $X$ failure probabilities then yields
%\begin{eqnarray}
%\frac{n}{2}\ln\left(4p_Xm\right) \approx \frac{m}{2}\ln\left(4p_Zn\right),
%\end{eqnarray}
%and hence 
%\begin{eqnarray}
%\frac{m}{n}\approx \frac{\ln\left(4p_Zm/b\right)}{\ln\left(4p_Zn\right)}.
%\end{eqnarray}
%The asymptotic behavior depends on the relative size of the number of qubits $q$ and the bias $b$. [{\em Blah, blah, blah, ...}]

\section{Comparison with numerics}
To check the accuracy of our formulas, we have also numerically determined the optimal values of $m$ and $n$, and the corresponding optimal failure probability, for a variety of values of $p_Z$ and $b$.

\begin{figure}[H]
\begin{center}
\epsfig{file=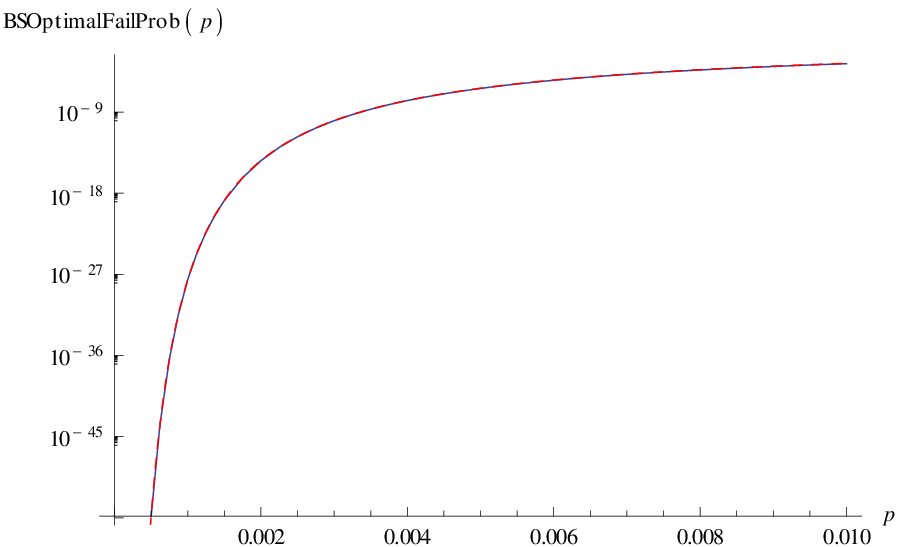,width=8cm} \vspace{0.2cm}
\end{center}
\fcaption{\label{fig:unbiased} The optimal failure probability for the Bacon-Shor code (solid blue curve) and the estimate Eq.(\ref{eq:symmetric-optimal}) (dashed red curve) for unbiased noise, as a function of $p\equiv p_Z = p_X$. The curves nearly coincide.}
\end{figure}

Fig.~\ref{fig:unbiased} shows the numerically optimized failure probability and the asymptotic estimate Eq.(\ref{eq:symmetric-optimal}) as a function of $p\equiv p_Z=p_X$ for the unbiased case ($b=1$). Our formula overestimates the exact result by less than 1\% for $p < .001$ and by less than 10\% for $p < .01$.

%Over the range plotted ($.0005 < p < .010$) the numerical result and the formula agree to within a few percent.

\begin{figure}[h!]
\begin{center}
\epsfig{file=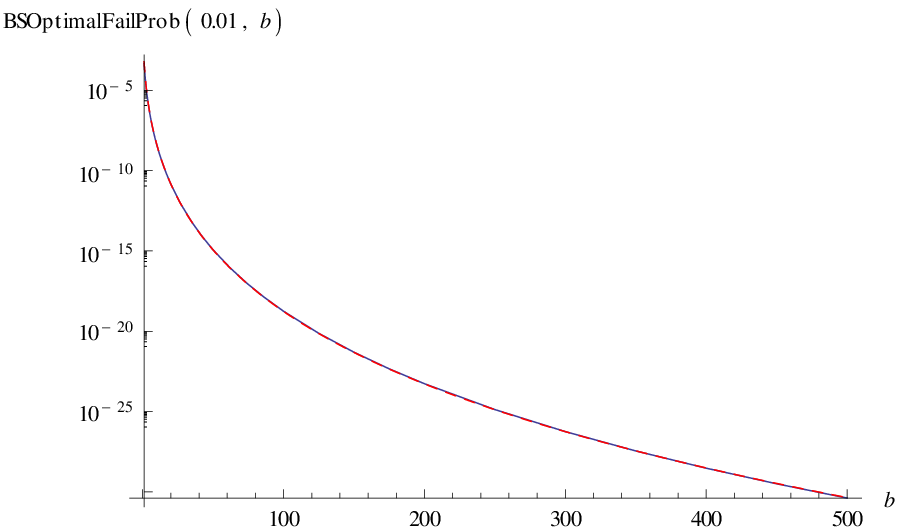,width=8cm} \vspace{0.2cm}
\end{center}
\fcaption{\label{fig:01} The optimal failure probability for the Bacon-Shor code (solid blue curve) and the estimate Eq.(\ref{eq:including-prefactors}) (dashed red curve) for $p_Z=.01$, as a function of the bias $b = p_Z/p_X$. The curves nearly coincide.}
\end{figure}

Fig.~\ref{fig:01} shows the numerically optimized failure probability and the estimate Eq.(\ref{eq:including-prefactors}) as a function of the bias $b$ for $p_Z = .01$. Here we have actually plotted the total error probability assuming $X$ and $Z$ errors are independent; that is,
\begin{eqnarray}
BSOptimalFailProb &=& BSOptimalZFailProb + BSOptimalXFailProb \nonumber\\
&&\quad - BSOptimalZFailProb \times BSOptimalXFailProb.
\end{eqnarray}
The agreement is good for $b > 10$, though the discrepancy grows with increasing $b$, reaching about 10\% for $b=500$.

\begin{figure}[h]
\begin{center}
\epsfig{file=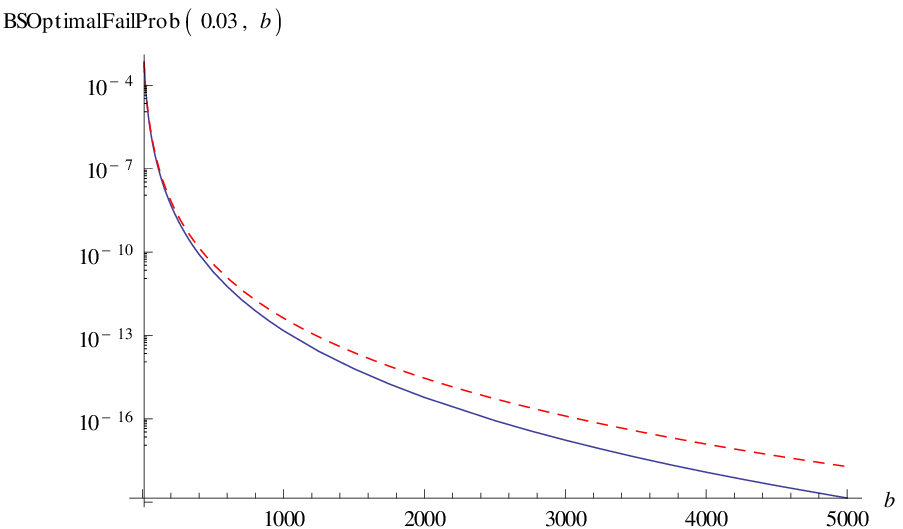,width=8cm} \vspace{0.2cm}
\end{center}
\fcaption{\label{fig:03} The optimal failure probability for the Bacon-Shor code (solid blue curve) and the estimate Eq.(\ref{eq:including-prefactors}) (dashed red curve) for $p_Z=.03$, as a function of the bias $b = p_Z/p_X$.}
\end{figure}

Fig.~\ref{fig:03} shows the numerically optimized failure probability and the estimate Eq.(\ref{eq:including-prefactors}) as a function of the bias $b$ for $p_Z = .03$. Now our formula badly overestimates the failure probability for large $b$, with the discrepancy reaching a factor of 13 for $b=5000$. In this regime, the condition $(.01) p_Z \ln^4 b\ll 1$ is not well satisfied ($(.01)p_Z\ln^4 b \approx 1.6$ for $b=5000$).

\begin{figure}[h]
\begin{center}
\epsfig{file=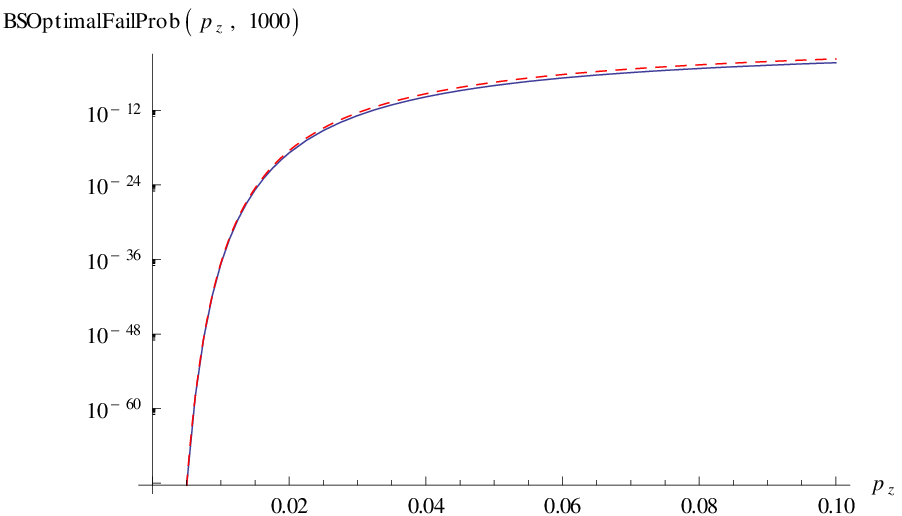,width=8cm} \vspace{0.2cm}
\end{center}
\fcaption{\label{fig:1000} The optimal failure probability for the Bacon-Shor code (solid blue curve) and the estimate Eq.(\ref{eq:including-prefactors}) (dashed red curve) for $b=1000$, as a function of the error probability $p_Z$.}
\end{figure}

\begin{figure}[H]
\begin{center}
\epsfig{file=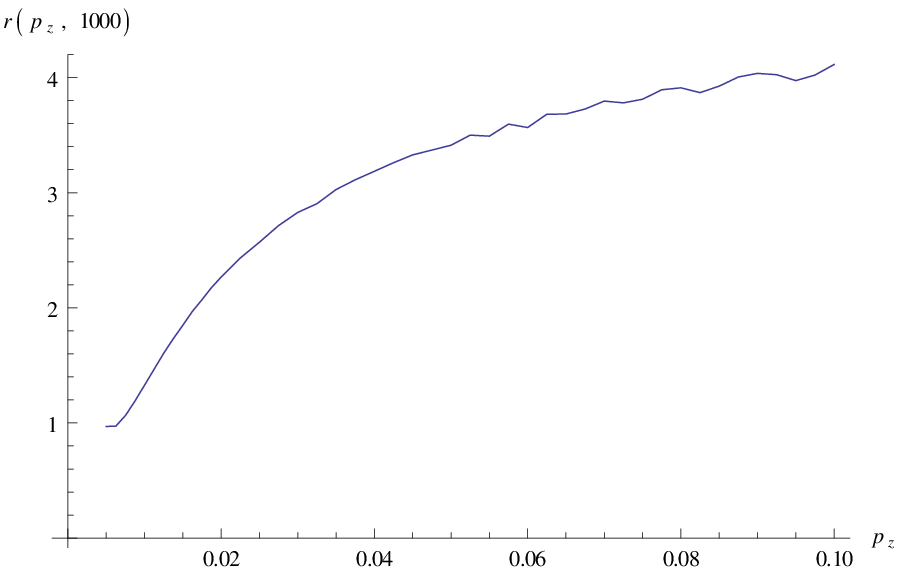,width=8cm} \vspace{0.2cm}
\end{center}
\fcaption{\label{fig:ratio-1000} The ratio of the optimal failure probability to the estimate Eq.(\ref{eq:including-prefactors}) for $b=1000$, as a function of the error probability $p_Z$. The kinks in the plot occur because the optimal dimensions $m\times n$ of the Bacon-Shor code are integers which change discontinuously.}
\end{figure}

Fig.~\ref{fig:1000} shows the numerical and analytic results as a function of $p_Z$ for $b=1000$, and Fig.~\ref{fig:ratio-1000} shows the ratio of the two, illustrating that the agreement improves rapidly as $p$ decreases. 

\section{2D Bacon-Shor recovery with syndrome measurement errors}

Measurement of the weight-two gauge operator $XX$ can be executed by a simple circuit with four locations --- e.g. preparation of an ancilla qubit in the $X=1$ eigenstate $|+\rangle$, two successive controlled-$X$ gates with ancilla qubit as control and data qubit as target, followed by measurement of the ancilla qubit in the $X$ basis. A similar circuit measures $ZZ$, but with the controlled-$X$ gates replaced by controlled-$Z$ gates. 

A fault in the circuit that flips the ancilla qubit could cause the measurement outcome to be incorrectly recorded, and a faulty two-qubit gate could damage a qubit in the code block. However, a single fault in the circuit will not cause two errors in the code block that cannot be gauged away. A $Z$-type syndrome bit is obtained by computing the parity of $n$ outcomes of $XX$ gauge qubit measurements, and an error in any one of those $n$ measurements could flip the bit, so that the probability of an error in a syndrome bit is roughly $n$ times larger than the probability of error in each gauge qubit measurement. Similarly, an $X$-type syndrome bit is obtained by computing the parity of $m$ outcomes of $ZZ$ gauge qubit measurements, so the probability of an error in a syndrome bit is roughly $m$ times the probability of error in each gauge qubit measurement. 

Though measurement and data errors may actually be correlated, these considerations motivate an idealized model for Bacon-Shor error recovery with noisy measurements. For the $Z$-type syndrome we consider a repetition code of length $m$, where both the probability of error per data bit and the probability of error per syndrome bit is $\tilde p_Z =OddProb(p_Z,n)$, and $p_Z$ is the fixed physical error rate for a $Z$-type error. Likewise, for the $X$-type syndrome we consider a repetition code of length $n$, where both the probability of error per data bit and the probability of error per syndrome bit is $\tilde p_X =OddProb(p_X,m)$, and $p_X$ is the fixed physical error rate for an $X$-type error.

We will study this model for both unbiased noise and highly biased noise with $b = p_Z/p_X \gg 1$. Justifying highly biased noise models raises subtle issues --- why should the nontrivial quantum gates used in the syndrome measurement circuit strongly favor $Z$ errors over $X$ errors \cite{Aliferis2008,Brooks2012}? We will not address these issues here; rather we shall be satisfied to say that a model in which both the data qubit errors and the syndrome bit errors are highly biased is mathematically natural and worthy of investigation. Ref. \cite{Brooks2012} contains a much more complete discussion of fault-tolerant error correction for Bacon-Shor codes.

A nearly optimal recovery scheme for a repetition code with syndrome errors was described in \cite{Dennis2002,Wang2003}. To obtain more reliable syndrome information we follow the history of the syndrome through many measurement cycles, and assume that  the probabilities for both data errors and syndrome errors are given by  $\tilde p_Z$, $\tilde p_X$ in each round of syndrome measurement. The syndrome history can be represented on a two-dimensional square lattice in spacetime, where each horizontal row of squares records the results from one cycle of (possibly noisy) syndrome measurement and each square represents a data bit --- a marked vertically oriented link in the row indicates that the two data bits sharing that link were found to have opposite parity. We identify all the boundary points of this chain of nontrivial syndrome bits, and apply the Edmonds matching algorithm to find the minimum-weight chain in spacetime with these boundary points (relative to the boundary of the sample). This minimum-weight chain identifies the most likely error history compatible with the observed syndrome history, where vertical edges in the chain are hypothetical syndrome measurement errors and horizontal edges are hypothetical data errors. The actual error chain combined with the hypothetical chain comprises a closed chain relative to the boundary, and a segment of this chain stretching across the sample, connecting it's left and right boundary, signifies a logical error. If we consider the repetition code on a circle rather than an open one-dimensional lattice, so there is no sample boundary, then a closed chain wrapping around the cylinder indicates a logical error.

This scheme has been studied previously \cite{Dennis2002,Wang2003} in the case where the probability $p$ of a data error or syndrome error is a constant independent of the length $m$ of the repetition code; in that case recovery is successful with a probability approaching 1 as $m\to \infty$ for $p < 10.3\%$ using the matching algorithm, and for $p < 11.0\% $ using the optimal algorithm. Now we wish to reconsider the efficacy of the matching algorithm in the case where the error rate scales nontrivially (roughly linearly) with $m$.

Based on our earlier results for the case of ideal syndrome measurement, we anticipate that if $m$ is chosen optimally then the probability per unit time of an encoded error in the Bacon-Shor code has the form
\begin{eqnarray}
\ln[BSOptimalFailRate(p_Z,b)]&\approx& -\tilde A(b)/p_Z +\tilde B(b)\ln p_Z +\tilde C(b) +\cdots,
\end{eqnarray}
but with different functions of the noise bias $\tilde A(b), \tilde B(b), \tilde C(b)$ than in the case of ideal measurements. We could try to estimate these functions using a Monte Carlo method, in which we generate sample error histories, infer the syndrome, find the minimum weight matching of the syndrome's boundary points, and then determine whether a logical error has occurred. This method is difficult to carry out, however, because for small $p_Z$ the optimal failure probability is quite small, and we need to generate many samples to estimate it with reasonable statistical accuracy. Instead, we will use an analytic argument to obtain a rather loose upper bound on $\tilde A$ (which dominates the scaling of the failure probability when $p_Z$ is small) for both the case of unbiased noise $(b \equiv p_Z/p_Z =1)$, and the case of highly biased noise $(b\equiv p_Z/p_X \gg 1)$. 

Since we expect the value of $\tilde A$ to be the same for a planar Bacon-Shor code as for a code defined on a torus, we will consider the case of a torus for convenience. Thus we consider the syndrome history (for either the $X$ or $Z$ errors) on a cylinder, closed in the spatial direction but open in the temporal direction. 
If there is a logical error, then the combination of the actual error chain and the hypothetical error chain must contain a self-avoiding cycle that wraps once around the cylinder, where at least half of the edges in this cycle have actual errors (otherwise there would a lower-weight choice for the hypothetical chain). If we fix a cycle of length $r$, then the probability that a particular set of $s$ edges in the cycle have actual errors, while the remaining $r-s$ edge do not, is 
\begin{equation}
\tilde p^s(1-\tilde p)^{r-s} = [\tilde p(1-\tilde p)]^{r/2} [\tilde p/(1-\tilde p)]^{s-r/2} \le [\tilde p(1-\tilde p)]^{r/2}
\end{equation}
where the inequality holds provided $\tilde p\le 1-\tilde p$ and $s\ge r/2$. The number ${r\choose s}$ of ways to place $m$ errors on the cycle is bounded above by $2^r$, therefore, the probability of failure arising from this particular length $r$ cycle is bounded above by $[4\tilde p(1-\tilde p)]^{r/2}$. 

Considering the $Z$ error syndrome history for definiteness, we now let $SAC(r,m)$ denote the number of distinct self-avoiding cycles of length $r$ that wrap around the cylinder of circumference $m$. Here we are not counting the freedom to translate the cycle in either space or time. Taking this freedom into account, we obtain an upper bound on the probability of error per unit time
\begin{eqnarray}
BSZFailRate(p_Z,m,n)\le m\sum_{r=m}^\infty SAC(r,m)[4\tilde p_Z(1-\tilde p_Z)]^{r/2}.
\end{eqnarray}
The factor of $m$ in front arises from the $m$ possible spatial translations of the cycle, and a factor of time $T$ arising from time translations has been divided out to obtain a failure rate per unit time. Note that the cycle length is in principle unbounded, as it could extend indefinitely in the time direction. 

%To go further we must estimate $SAC(r,L)$. Considerable computational effort has been devoted to estimating the number of self-avoiding polygons in the plane (see for example Appendix C of {\em The Self-Avoiding Walk} by Madras and Slade, or arXiv:cond-mat/9905291 by Jensen and Gutmann) but I am not aware of comparable studies of the number of cycles on the cylinder or torus. From the data in these references, the number $SAP(r)$ of length-$r$ self-avoiding polygons in the plane seems to fit 
%\begin{equation}
%SAP(r) \approx (.5569)r^{-2.5}(2.63816)^r,
%\end{equation}

The number $SAC(r,m)$ of self-avoiding cycles of length $r$ on the cylinder is bounded above by the number $SAW(r)$ of self-avoiding open walks with a specified starting point, for which an upper bound is known of the form \cite{Madras1996,Jensen2004}
\begin{equation}
SAW(r) \le \gamma r^{\beta}\mu^r,
\end{equation}
where $\mu \approx 2.6381585$, $\beta = 11/32$, and $\gamma\approx 1.17704$. Plugging the upper bound on $SAW(r)$ into our expression for $BSZFailRate$, we find 
\begin{eqnarray}\label{eq:noisyZandX}
BSZFailRate(p_Z,m,n)&\le& \gamma m\sum_{r=m}^\infty r^\beta \mu^r[4\tilde p_Z(1-\tilde p_Z)]^{r/2}\nonumber\\
 &\le& \gamma m m^\beta [4\mu^2\tilde p_Z(1-\tilde p_Z)]^{m/2} \sum_{s=m}^\infty \left(\frac{m+s}{m}\right)^\beta [4\mu^2\tilde p_Z(1-\tilde p_Z)]^{s/2}\nonumber\\
&\le& \gamma m^{\beta+1} [4\mu^2\tilde p_Z(1-\tilde p_Z)]^{m/2} \sum_{s=0}^\infty e^{\beta s/m} [4\mu^2\tilde p_Z(1-\tilde p_Z)]^{s/2}\nonumber\\
&=& \gamma m^{\beta+1} [4\mu^2\tilde p_Z(1-\tilde p_Z)]^{m/2} \frac{1}{1-\sqrt{4\mu^2 e^{2\beta/m}\tilde p_Z (1-\tilde p_Z)}}.
\end{eqnarray}
We will obtain our upper bound on the failure rate by choosing $m$ and $n$ such that $4\mu^2 e^{2\beta/m}\tilde p_Z(1-\tilde p_Z)$ is strictly less than one; therefore we can bound the last factor by a constant, obtaining
\begin{eqnarray}
\ln[BSZFailRate(p_Z,m,n)]\le \frac{m}{2}\ln[4\mu^2\tilde p_Z(1-\tilde p_Z)] + O(\ln m), 
\end{eqnarray}
and by similar reasoning
\begin{eqnarray}
\ln[BSXFailRate(p_X,m,n)]\le \frac{n}{2}\ln[4\mu^2\tilde p_X(1-\tilde p_X)] + O(\ln n). 
\end{eqnarray}

\subsection{Unbiased noise}
In the case of unbiased noise, where $p_Z=p_X\equiv p$, we choose $n=m$ and recall that $\tilde p = \frac{1}{2}\left( 1 - (1-2p)^m\right) \approx \frac{1}{2}\left( 1- e^{-2pm}\right)$, so that 
\begin{equation}
4\tilde p(1-\tilde p) \approx 1 - e^{-4pm}.
\end{equation}
Thus we conclude that the failure rate for either $Z$-type or $X$-type errors is
\begin{eqnarray}
\ln[BSFailRate(p,m)]\le \frac{m}{2}\ln\left[\mu^2\left(1-e^{-4pm}\right)\right] + O(\ln m). 
\end{eqnarray}

To find the value of $m$ that minimizes this upper bound (for asymptotically large $m$), we solve
\begin{equation}
0= \frac{1}{2}\left(\ln\left[\mu^2\left(1-e^{-4pm}\right)\right] + \frac{4pm}{e^{4pm}-1}\right) + O(1/m),
\end{equation}
for $\mu = 2.6381585$, finding $pm\approx 0.0139682$, so that
\begin{eqnarray}
\ln[BSOptimalFailRate(p)] \le -\frac{.00679079}{p} + O(\ln p).
\end{eqnarray}
Thus we derive a lower bound
\begin{equation}\tilde A \ge .00679079,
\end{equation}
which is about an order of magnitude smaller than the value of $A$ found for the case of ideal syndrome measurement.

Even more simply, we may observe that $\tilde p \le pm$ and $\tilde p(1-\tilde p)\le \tilde p$ to obtain
\begin{eqnarray}
\ln[BSFailRate(p,m)]\le \frac{m}{2}\ln(4\mu^2 pm) + O(\ln m),
\end{eqnarray}
which is optimized by choosing $4\mu^2 pm = 1/e$, so that 
\begin{eqnarray}
\ln[BSOptimalFailRate(p)] \le -\frac{1}{8\mu^2 e p} +O(\ln p)  = -\frac{.00660714}{p} +O(\ln p),
\end{eqnarray}
which provides nearly as good a lower bound on $\tilde A$. 

\subsection{Highly biased noise}
In the case of biased noise, again using the approximations $\tilde p_Z(1-\tilde p_Z)\le p_Zn$ and $\tilde p_X(1-\tilde p_X)\le p_Xm$, the leading contributions to the $Z$ and $X$ failure rates are
\begin{eqnarray}\label{eq:approx-noisy-failure-rate}
\ln[BSZFailRate(p_Z, m, n)]\approx \frac{m}{2}\ln(4\mu^2p_Zn)&=& \frac{1}{8\mu^2p_X}X\ln Z=\frac{1}{8\mu^2p_Z}bX\ln Z, \nonumber\\
\ln[BSXFailRate(p_X, m, n)]\approx \frac{n}{2}\ln(4\mu^2p_Xm)&=&\frac{1}{8\mu^2p_Z}Z\ln X,
\end{eqnarray}
where
\begin{eqnarray}
X = 4\mu^2 p_X m, \quad Z= 4\mu^2p_Z n, \quad b= \frac{p_Z}{p_X}.
\end{eqnarray}
(Actually, the approximation $p_Z(1-\tilde p_Z)\approx p_Zn$ is not necessarily so accurate when $m$ and $n$ are chosen optimally, but it suffices for deriving an upper bound on the failure rate; if we do not make this approximation the analysis becomes more complicated and the upper bound improves by only about 8\%.)
Equating the upper bounds on the $X$ and $Z$ failure rates, we are to minimize
\begin{eqnarray}
F(X,Z) = Z\ln X
\end{eqnarray}
subject to the constraint
\begin{eqnarray}\label{eq:noisy-constraint}
Z\ln X = bX \ln Z.
\end{eqnarray}
Introducing a Lagrange multiplier $\lambda$ to do the constrained minimization, we obtain the equations
\begin{eqnarray}
\left(1+\lambda\right)\ln X &=& \lambda b \frac{X}{Z},\nonumber\\
\left(1+\lambda\right) \frac{Z}{X} &=& \lambda b \ln Z,
\end{eqnarray}
which imply
\begin{eqnarray}
\ln X = \frac{1}{\ln Z}.
\end{eqnarray}
Using the constraint Eq.(\ref{eq:noisy-constraint}), we see that the aspect ratio of the optimal code is 
\begin{eqnarray}
\frac{m}{n} = b\frac{X}{Z} = \frac{\ln X}{\ln Z} = \ln^2 X.
\end{eqnarray}

When the bias $b$ is very large, the optimal value of $Z$ (that is, when $m$ and $n$ are chosen to optimize the approximate expression Eq.(\ref{eq:approx-noisy-failure-rate}), an upper bound on the actual failure rate) approaches one from below, while $4\mu^2 e^{2\beta/m}\tilde p_Z(1-\tilde p_Z)$ remains strictly less than one for $m$ large, so we can still bound the last factor in Eq.(\ref{eq:noisyZandX})  by a constant. To estimate the failure probability, we first determine $Z$ by solving Eq.(\ref{eq:noisy-constraint}) in the form
\begin{eqnarray}\label{eq:noisy-Z}
Z = b~ \ln^2 Z ~\exp(1/\ln Z).
\end{eqnarray}
Then we find $X$ by substituting this value of $Z$ back into Eq.(\ref{eq:noisy-constraint}), finding
\begin{eqnarray}
\frac{b}{Z} = \frac{\ln^2 X}{X} \Rightarrow \frac{1}{\sqrt{X}} \ln\left(\frac{1}{\sqrt{X}}\right) = \sqrt{\frac{b}{4Z}}.
\end{eqnarray}
Recalling that the Lambert $W$ function $W(z)$ is defined by $We^W = z$, we conclude that
\begin{eqnarray}
\ln\left(\frac{1}{\sqrt{X}}\right)= W\left(\sqrt{\frac{b}{4Z}}\right).
\end{eqnarray}
Thus we find that the value of $F$ at its minimum is
\begin{eqnarray}
F_{\rm min} = Z \ln X \approx -2 Z~W\left(\sqrt{\frac{b}{4Z}}\right),
\end{eqnarray}
and we obtain upper bounds on the leading contributions to the failure rates:
\begin{eqnarray}
\ln[BSZFailRate(p_Z, m, n)]\approx \ln[BSXFailRate(p_X, m, n)]\approx -\frac{Z(b)}{4\mu^2p_Z} W\left(\sqrt{\frac{b}{4Z(b)}}~\right),
\end{eqnarray}
where $Z(b)$ is a slowly varying monotonic function, ranging from $1/e$ to $1$ as $b$ increases from 1 to infinity.
Comparing to the upper bound for the case of ideal syndrome measurement we find
\begin{eqnarray}
A / \tilde A  = \frac{\mu^2}{2Z(b)} \frac{W^2(\sqrt{b})}{W(\sqrt{b/4Z})},
\end{eqnarray}
a slowly increasing function of $b$.
The optimal aspect ratio is 
\begin{eqnarray}
\frac{m}{n}= \ln^2 X \approx 4 W^2\left(\sqrt{\frac{b}{4Z(b)}}~\right);
\end{eqnarray}
in contrast to the case where the syndrome is perfect, this aspect ratio grows only polylogarithmically with the bias $b$.

For example, by solving Eq.(\ref{eq:noisy-Z}), we find $Z=.3679$, $.7046$, $.8428$, $.8994$ for $b=1$, $10^2$, $10^4$, $10^6$ and corresponding values of $-F_{\rm min}=.3679 (=1/e)$, $2.01255$, $4.92966$, $8.48426$.
For $b=10^2$, $10^4$, $10^6$, our values of  of $A/\tilde A$ are $5.231$,  $11.50$, and $18.29$ respectively.

\section{Higher-dimensional Bacon-Shor codes}

The two-dimensional Bacon-Shor code can be extended to higher dimensions. Bacon \cite{Bacon2006} discussed one such extension to three dimensions. In Bacon's code, defined on an $L \times L\times L$ cubic lattice (for $L$ odd), the logical operators $\bar X$ and $\bar Z$ are both weight $L^2$ operators defined on orthogonal planes, and a length $L$ repetition code protects against both $X$ and $Z$ errors. Here we will briefly discuss a different way to extend the code to three (or more) dimensions, in which the logical operator $\bar Z$ resides on a horizontal plane and the logical operator $\bar X$ resides on a vertical line. This alternative formulation is less symmetric but in some ways more natural; it may be more effective than Bacon's code if the noise is highly biased, and it is more robust against errors in the measurement of the $Z$-type error syndrome

We consider an $m\times n\times k$ cubic lattice, and choose the gauge algebra to be generated by two-qubit operators $XX$ acting on pairs of neighboring qubits in the same horizontal plane, and by two-qubit operators $ZZ$ acting on pairs of qubits in the same vertical column. There is one protected qubit; the weight-$mn$ operator $\bar Z$ acts on all qubits in a horizontal plane and the weight-$k$ operator $\bar X$ acts on all the qubits in a vertical column. The stabilizer group has $mn-1$ independent $X$-type generators, each a product of $2k$ $X$s acting on all the qubits in a pair of neighboring vertical columns, and $k-1$ independent $Z$-type generators, each a product of $2mn$ $Z$s acting on all the qubits in a pair of neighboring horizontal planes. As in the 2D code, the value of a check operator can be found by measuring weight-two gauge operators and combining the results. For example, the weight-$2mn$ product of $Z$s in two neighboring planes is obtained by measuring $mn$ gauge operators $ZZ$, each acting on two qubits in the same column, and multiplying the results together. 

%We can check the counting of independent commuting Pauli operators as follows. In each of $k$ planes, there are $mn -1$ independent $XX$ gauge operators. In addition, there are $k-1$ $Z$-type stabilizer generators that commute with these, plus one logical Pauli operator, for a total of 
%\begin{equation}
%k(mn-1) + k = mnk
%\end{equation}
%independent mutually commuting operators on the $mnk$ qubits. Alternatively, there are $k-1$ independent $ZZ$ gauge operators in each of $mn$ columns, plus $mn-1$ $X$-type stabilizer generators and one logical Pauli operator, for a total of 
%\begin{equation}
%(k-1)mn + mn = mnk
%\end{equation}
%independent mutually commuting operators. 

As in the 2D case, the error recovery procedure is clarified if we adopt a suitable gauge. An even number of $Z$ errors in a column can be gauged away, and if there are an odd number of $Z$ errors in a column, there is a gauge such that all errors but one are removed, and the single error lies in the uppermost horizontal plane. Similarly, an even number of $X$ errors in a plane can be gauged away, and if there are an odd number of $X$ errors in a plane, all but one can be removed, with the single error lying in the northwest corner of the plane. The check operators, whose value can be inferred from measurements of many weight-two gauge operators, check whether qubits in the northwest corner of neighboring planes agree in the $Z$ basis, and whether qubits in the top plane of neighboring columns agree in the $X$ basis. 

The planar repetition code corrects $(mn -1)/2$ $Z$ errors in the plane, and the columnar repetition code corrects $(k-1)/2$ $X$ errors in the column. Aside from its greater length (assuming $mn > k$), the planar code has another advantage over the columnar code --- measuring all local $X$-type gauge operators determines the error syndrome redundantly, which improves the robustness against syndrome measurement errors. From the gauge qubit measurements, we determine the value of $X^{\otimes 2k}$ for any two neighboring columns; there are $m(n-1)$ such pairs of ``east-west'' neighbors as well as $n(m-1)$ pairs of ``north-south'' neighbors, for a total of $2mn - m - n$ check bits, or which only $mn-1$ are independent. 

Recovery from $Z$ errors succeeds if the number of vertical columns with an odd number of $Z$ errors is no more than $(mn-1)/2$. and recovery from $X$ errors succeeds if the number of horizontal planes with an odd number of $X$ errors is no more than $(k-1)/2$. Setting $m=n$, therefore, the probability of a logical $Z$ error in an independent error model (assuming perfect syndrome measurements) is
\begin{eqnarray}
BS3DZFailProb(p_Z,m,k)= RepFailProb(OddProb(p_Z,k),m^2)
\end{eqnarray} 
where $p_Z$ is the $Z$ error rate, and the probability of a logical $X$ error is
\begin{eqnarray}
BS3DXFailProb(p_X,m,k)= RepFailProb(OddProb(p_X,m^2),k)
\end{eqnarray} 
where $p_X$ is the $X$ error rate. These are the same functions we encountered in our analysis of the 2D code, except with $m$ replaced by $m^2$ and $n$ by $k$.

In four dimensions, denote the four directions by $x,y,z,w$. To define the Bacon-Shor code, choose the gauge algebra to be generated by $XX$ acting on neighboring qubits in each $xy$-plane (with $z$ and $w$ fixed) and by $ZZ$ acting on neighboring qubits in each $zw$-plane (with $x$ and $y$ fixed). Note that each $xy$-plane intersects exactly once with each $zw$-plane. The logical $\bar Z$ is the product of all $Z$s in an $xy$-plane, and the logical $\bar X$ is the product of all $X$s in a $zw$-plane. For a suitable gauge choice, all $X$ errors can be moved to the $zw$-plane with $x=y=0$ and all $Z$ errors can be moved to the $xy$-plane with $z=w=0$. Hence for this four-dimensional Bacon-Shor code a two-dimensional repetition code protects against logical $Z$ errors and another two-dimensional repetition code protects against logical $X$ errors.  If we choose the dimensions of the hypercubic lattice to be $m\times m\times n\times n$, then the probability of a logical $Z$ error (assuming perfect syndrome measurement) in an independent noise model is 
\begin{eqnarray}
BS4DZFailProb(p_Z,m,n)= RepFailProb(OddProb(p_Z,n^2),m^2)
\end{eqnarray} 
where $p_Z$ is the physical $Z$ error probability per qubit, and the probability of a logical $X$ error is
\begin{eqnarray}
BS4DXFailProb(p_X,m,n)= RepFailProb(OddProb(p_X,m^2),n^2),
\end{eqnarray} 
where $p_X$ is the physical $X$ error probability per qubit.
These are the same functions as for the 2D code, but with $m$ replaced by $m^2$ and $n$ by $n^2$.
The advantage of the four-dimensional code over the two-dimensional or three-dimensional code is that measuring the weight-two local gauge operators provides redundant syndrome information (and hence improved robustness against syndrome measurement errors) for both the $Z$ and $X$ errors.

\section{Conclusions}

We have studied the performance of Bacon-Shor codes against both unbiased and biased noise, concluding that the codes provide excellent protection against logical errors if the error probability per qubit is less than a few tenths of a percent in the unbiased case, and if the dephasing error probability is less than a few percent in the case of highly biased noise, assuming the syndrome is measured perfectly.

Using quantum codes to protect quantum computers from noise raises many thorny problems \cite{Shor1996,Gottesman2009}; in particular, we need to build a set of gadgets that reliably execute logical gates acting on the code space, and in the case of biased noise we need to employ a physical gate set compatible with the bias \cite{Aliferis2008,Brooks2012}. We have not discussed these issues here. Instead, our goal has been to gain a better understanding of the properties of the Bacon-Shor code family, without getting bogged down in detailed constructions of fault-tolerant protocols. We expect, though, that our analytic formulas for the logical failure probabilities of optimal codes will provide helpful guidance for the construction of fault-tolerant schemes. 

\nonumsection{Acknowledgments}
\noindent
We thank Peter Brooks and Franz Sauer for valuable discussions.
%%%
This work was supported in part by the Intelligence Advanced Research Projects Activity (IARPA) via Department of Interior National Business Center contract number D11PC20165. The U.S. Government is authorized to reproduce and distribute reprints for Governmental purposes notwithstanding any copyright annotation thereon. The views and conclusions contained herein are those of the author and should not be interpreted as necessarily representing the official policies or endorsements, either expressed or implied, of IARPA, DoI/NBC
or the U.S. Government.
%%%
We also acknowledge support from NSF grant PHY-0803371, DOE grant DE-FG03-92-ER40701, NSA/ARO grant W911NF-09-1-0442, Caltech's Summer Undergraduate Research Fellowship (SURF) program, and the Victor Neher SURF Endowment. The Institute for Quantum Information and Matter (IQIM) is an NSF Physics Frontiers Center with support from the Gordon and Betty Moore Foundation. 

\nonumsection{References}

\end{document}